\documentclass{aa}
\usepackage{graphicx}
\usepackage{txfonts}
\begin{document}
   \title{Angular momentum transport and element mixing in the stellar interior}

   \subtitle{I. Application to the rotating Sun}

   \author{Yang, W. M.
          \inst{1, 2}
          \and
          Bi, S. L.
          \inst{1}
          }

   \offprints{WuMing Yang}

   \institute{National Astronomical Observatories/Yunnan Observatory, Chinese
              Academy of Sciences,
             \\Kunming 650011, China\\
             \email{yangwuming@ynao.ac.cn}\\
             \email{bislan@public.km.yn.cn}
             \and
             Graduate School of The Chinese Academy of Sciences,
             Beijing 100039, China\\
             }

   \abstract
    {}
    {The purpose of this work was to obtain diffusion coefficient for
    the magnetic angular momentum transport and material transport in
    a rotating solar model.}
    {We assumed that the transport of both angular momentum and chemical
    elements caused by magnetic fields could be treated as a diffusion
    process. }
    {The diffusion coefficient depends on the stellar
    radius, angular velocity, and the configuration of magnetic fields. By
    using of this coefficient, it is found that our model becomes more
    consistent with the helioseismic results of total angular
    momentum, angular momentum density, and the rotation rate in a radiative
    region than the one without magnetic fields. Not only can the magnetic
    fields redistribute angular momentum efficiently, but they can
    also strengthen the coupling between the radiative and convective zones.
    As a result, the sharp gradient of the rotation rate is reduced at the
    bottom of the convective zone. The thickness of the layer of sharp radial
    change in the rotation rate is about 0.036 $R_{\odot}$ in our model.
    Furthermore, the difference of the sound-speed square between the seismic
    Sun and the model is improved by mixing the material that is associated
    with angular momentum transport.}
    {}

   \authorrunning{Yang, W. M. \& Bi, S. L. }
   \titlerunning{Angular momentum transport and element mixing}
   \maketitle

%
%________________________________________________________________

\section{Introduction}

   In standard stellar models, it is assumed that there are no rotation and
   magnetic fields. Although this framework is strongly supported by
   helioseismology, recent advances in the study of solar structure shows
   that differences exist between the Sun and this model. Those differences
   are small and display a very interesting systematic behavior, which is
   far from satisfactory (Christensen-Dalsgaard  \cite{Christensen02}). The
   most striking difference is the bump in the sound speed just beneath the
   convection zone. Noting both $c^2\propto T/\mu$ and the $\mu$-gradient
   caused by microscopic diffusion and element settling in the radiative
   region, Gough et al. (\cite{Gough96}) and Christensen-Dalsgaard et al.
   (\cite{Christensen96}; \cite{Christensen02}) argued that material
   mixing may be important but has been neglected.

   Helioseimology also shows that rotation is almost uniform in the solar
   radiative region: the Sun has a slow rotation core, the angular velocity
   $\Omega$ has a small radial gradient and large latitudinal gradient in the
   convective zone, and the radial angular velocity gradient
   is positive at low latitudes and negative at high latitudes in the tachocline
   (Schou et al. \cite{Schou}; Chaplin et al. \cite{Chaplin}). But the
   calculations of rotational models show that the sun has a fast rotation
   core and a large gradient of the rotation rate in the radiative region
   (Pinsonneault et al. \cite{Pinsonneault}; Charboyer et al. \cite{Chaboyer95}),
   which disagrees with the helioseismic results. Thompson
   et al. (\cite{Thompson}) argues that some mechanisms of angular
   momentum transport may be missed in these models. Magnetic and gravity waves
   mechanisms have been proposed; however, what mechanisms are causing the
   nearly uniform rotation in the radiative region are poorly understood.

   Rotation is a property that virtually all stars possess.
   Rotation affects stellar structure mainly in two ways. The first
   is an immediately dynamical effect caused by centrifugal acceleration
   of the hydrostatic balance. In most cases, however, this effect is
   modest and can be included through modifying the equations
   of stellar structure. Although the equations of stellar structure
   of a rotating star are three-dimensional, the method of Kippenhahn \&
   Thomas (\cite{Kippenhahn70}), which uses the mass contained within an
   equipotential surface, $M_{\Psi}$, as an independent variable, allows
   a one-dimensional evolution code to be modified to incorporate
   the hydrostatic effects of rotation. This method was modified by
   Endal \& Sofia (\cite{Endal76}) and Meynet \& Maeder (\cite{ Meynet})
   to apply to shellular rotation (Zahn \cite{Zahn92}), a rotation
   rate that depends only on the radial coordinate $r$ to a first approximation.
   The other way is an effect arising from the redistribution
   of chemical elements due to the instabilities caused by rotation. This effect
   is much more important and it has been studied by many investigators (Endal \&
   Sofia \cite{Endal78}; Pinsonneault et al. \cite{Pinsonneault}; Chaboyer \&
   Zahn \cite{Chaboyer92}; Zahn \cite{Zahn92}; Meynet \& Maeder \cite{Meynet};
   Maeder \cite{Maeder97}; Maeder \& Zahn \cite{Maeder98}; Maeder \&
   Meynet \cite{Maeder00}; Huang \cite{Huang04a}, \cite{Huang04b}). The method
   of Kippenhahn \& Thomas (\cite{Kippenhahn70}) as
   modified by Meynet \& Maeder (\cite{Meynet}) is used in our model.

   Magnetic fields are another property of stars. Their effects on a star
   create a complicated problem. The magnetic field is involved
   in most problems in astrophysics. But the generation of magnetic fields
   is still a controversial problem. They may be fossil fields
   (Cowling \cite{Cowling}; Moss \cite{Moss}; Braithwaite \cite{Braithwaite}),
   which are remnants of the star's formation, or may be generated by a
   convective stellar dynamo (Parker \cite{Parker}; Charbonneau \& MacGregor
   \cite{Charbonneau01}) in the convective zone or generated by a Tayler-Spruit
   dynamo (Pitts \& Tayler \cite{Pitts}; Spuit \cite{Spruit02}) in a
   differential rotation star. In this paper we assumed that the magnetic
   fields exist. We only consider the process of magnetic angular momentum
   transport and material mixing. Although the magnetic angular momentum
   transport has been treated by many investigators (Charbonneau \&
   MacGregor \cite{Charbonneau92}, \cite{Charbonneau93}; Spruit
   \cite{Spruit02}; Maeder \& Meynet \cite{Maeder03}, \cite{Maeder04}),
   it is still an open question.

   In this paper, we focus mainly on magnetic angular momentum transport and
   the mixing of elements associated with angular momentum redistribution. In
   Sect. 2 we give the equations of stellar structure of a shellular rotation
   star. In Sect. 3 we deduce the diffusion coefficient for magnetic angular
   momentum transport and material mixing that is due to angular momentum
   redistribution. Then, in Sect. 4 we give the results of the numerical
   calculation. We then discuss our results and conclude in Sect. 5.

%__________________________________________________________________
\section{Equations of stellar structure of a shellular rotation star}
     The stars whose angular velocity is constant on their isobars
     are called shellular rotation stars (Meynet \& Maeder
     \cite{Meynet}); the isobar surfaces are given by (Meynet \& Maeder
     \cite{Meynet})
     \begin{equation}
        \Psi_{p}=\Phi+\frac{1}{2}\Omega^{2}r^{2}sin^{2}\theta=constant\,,
      \label{isobar}
     \end{equation}
     where $\Phi$ is the minus gravitational potential, $\Omega$
     the angular velocity, $r$ the radius, and $\theta$ the
     colatitude. The area of such an isobar surface is denoted by
     $S_{p}$, and the volume enclosed by the isobar surface by
     $V_{p}$. For any quantity $q$, which is not constant over an
     isobar surface, a mean value is defined by
     \begin{equation}
      <q>=\frac{1}{S_{p}}\int_{\psi=const}qd\sigma\,,
     \end{equation}
     where $d\sigma$ is an element of the isobar surface.

    The equations of stellar structure of a shellular rotation star,
    which were developed by Kippenhahn \& Thomas (\cite{Kippenhahn70}) and
     Meynet \& Maeder (\cite{Meynet}), are as follows:
    \begin{equation}
       \frac{\partial P}{\partial M_{p}}=-\frac{GM_{p}}{4\pi
        r^{4}_{p}}f_{p}\,,
     \label{hyeq}
    \end{equation}
    \begin{equation}
       \frac{\partial r_{p}}{\partial M_{p}}=\frac{1}{4\pi
       r^{2}_{p}\bar{\rho}}\,,
     \label{maeq}
    \end{equation}
    \begin{equation}
       \frac{\partial L_{p}}{\partial M_{p}}=\epsilon_{n}-\epsilon_{\nu}+
       \epsilon_{g}\,,
     \label{eneq}
    \end{equation}
    \begin{equation}
       \frac{\partial T}{\partial M_{p}}=-\frac{GM_{p}}{4\pi
       r^{4}_{p}}f_{p}min[\nabla_{ad},\nabla_{rad}\frac{f_{T}}{f_{p}}]\,,
     \label{treq}
    \end{equation}
    where $r_{p}$ is the radius of a sphere enclosing the
    volume $V_{p}$, i.e.,
     \begin{equation}
       V_{p}=\frac{4\pi}{3}r_{p}^{3}\,.
      \label{volume}
     \end{equation}
     The $M_{p}$ is the mass inside the isobar,
    and
    \begin{equation}
       f_{p}=\frac{4\pi r^{4}_{p}}{GM_{p}S_{p}}\frac{1}{<g^{-1}_{e}>}\,,
    \end{equation}
    \begin{equation}
       \bar{\rho}=\frac{\rho(1-r^{2}sin^{2}\theta\Omega\alpha)<g^{-1}_{e}>}
       {<g^{-1}_{e}>-<g^{-1}_{e}r^{2}sin^{2}>\Omega\alpha}\,,
     \label{rhobar}
    \end{equation}
    \begin{equation}
       f_{T}=(\frac{4\pi
       r^{2}_{p}}{S_{p}})\frac{1}{<g_{e}><g^{-1}_{e}>}\,.
    \end{equation}
    Here $g_{e}$ is the effective gravity. The $\alpha$ in Eq. (\ref{rhobar})
    is a scalar $\frac{d\Omega}{d \Psi}$. Equation (\ref{hyeq}) is
    the hydrostatic equilibrium equation including centrifugal force,
    while Eq. (\ref{treq}) is the equation of energy transport under
    the effects of rotation. The nondimensional rotating corrective factors
    $f_{p}$ and $f_{T}$ depend on the shape of the isobars.

    Assuming that the shapes of isobars are spheroids with semi-major
    axis $a$ and semi-minor axis $b$, and given the angular velocity
    distribution, using definitions
    (\ref{volume}) and (\ref{isobar}), i.e.,
    \begin{equation}
       \frac{4}{3}\pi a^{2}b=\frac{4}{3}\pi r^{3}_{p}\,,
    \end{equation}
    \begin{equation}
       \frac{GM_{p}}{b}=\frac{GM_{p}}{a}+\frac{1}{2}\Omega^{2}a^{2}\,,
    \end{equation}
    we can get the values of $a$ and $b$ for any given $M_{p}$ and
    $r_{p}$; i.e., the surface of the isobar is determined.
    Thus, the average values of the effective gravity and its
    inverse ($<g_{e}>$ and $<g^{-1}_{e}>$) can be obtained. Then
    the $f_{p}$ and $f_{T}$ are also obtained. In the computation,
    the Roche mode (Kippenhahn \& Thomas \cite{Kippenhahn90}) was used
    to compute the gravitational potential.

\section{Angular momentum transport and the mixing of elements}

\subsection{Hydrodynamical instabilities}
    The mechanisms that redistribute angular momentum and the chemical elements
    in a rotating star can be divided into two categories according to the
    time scale involved.

    The first category is that of the dynamical
    instabilities, occurring on the dynamical timescale. If the dynamical
    unstable gradient occurs in star, it can be instantaneously smoothed
    (Endal \& Sofia \cite{Endal78}). One of the dynamical
    instabilities is the convective instability. We suppose that solid-body
    rotation was enforced in all convective regions. Another much more
    important is dynamical shear instabilities. These instabilities can ensure
    that the rotation velocity is constant on equipotential surfaces (
    Pinsonneault et al. \cite{Pinsonneault}).

    The second category, with a time scale that is comparable to the
    Kelvin-Helmholtz time scale or the time scale for the evolution of
    the star, is that of secular instabilities. For secular instabilities,
    the transport process of angular momentum and chemical composition
    was treated as a diffusion process (Endal \& Sofia \cite{Endal78};
    Pinsonneault et al. \cite{Pinsonneault}), so the radial equations for
    the redistribution of angular momentum and for the mass fraction
    $X_{i}$ are
    \begin{equation}
       \frac{\partial \Omega}{\partial t}=f_{\Omega}
       \frac{1}{\rho r^{4}}\frac{\partial}{\partial r}(\rho r^{4}D_{d}
       \frac{\partial \Omega}{\partial r})-\frac{1}{\rho r^{4}}
       \frac{\partial}{\partial r}(\rho r^{4}\Omega \dot{r}) \,,
      \label{diffu1}
    \end{equation}
    \begin{equation}
    \begin{array}{lll}
       \frac{\partial X_{i}}{\partial t}&=&f_{c}f_{\Omega}\frac{1}{\rho r^{2}}
       \frac{\partial}{\partial r}(\rho r^{2}D_{d}\frac{\partial X_{i}}
        {\partial r})\\
        & &+(\frac{\partial X_{i}}{\partial t})_{nuc}-\frac{1}
       {\rho r^{2}}\frac{\partial}{\partial r}(\rho r^{2}X_{i}V_{i}) \,,
     \end{array}
      \label{diffu2}
    \end{equation}
    where $D_{d}$ is the diffusion coefficient. Because of some inherent
    uncertainties in the diffusion equation, the adjustable parameter
    $f_{\Omega}$ is introduced to represent these uncertainties. Another
    adjustable parameter $f_{c}$ is used to account for how the
    instabilities mix material less efficiently than they transport angular
    momentum (Pinsonneault et al. \cite{Pinsonneault}). The second term
    on the right-side of Eq. (\ref{diffu1}) is due to the secular contraction
    and/or expansion (Maeder \& Zahn \cite{Maeder98}), which can be dominant to
    induce differential rotation in a rotating star with a weak wind or
    without any wind.
    The second term on the right-side of Eq. (\ref{diffu2}) is the change
    in the nuclear reaction. The $V_{i}$ in the Eq. (\ref{diffu2}) is the
    velocity of microscopic diffusion given by Thoul et al. (\cite{Thoul}).
    In our model, we use the diffusion coefficient that was given by
    Zahn (\cite{Zahn93}) for secular shear instability,
    \begin{equation}
       D_{d}=\frac{2c}{27G}|\frac{dln T}{dr}-\frac{2}{3}
       \frac{dln\rho}{dr}|^{-1}\frac{r^{4}}{\kappa \rho M(r)}
       (\frac{d\Omega}{dr})^{2} \,.
      \label{difcoe}
    \end{equation}

    It has been thought that the Sun and T-Tauri with about one solar mass
    both possess an initial angular momentum of the order of $10^{50}$ g $cm^{2}s^{-1}$
    (Kawaler \cite{Kawaler87}; Patern\`{o} \cite{Paterno}). The present total solar
    angular momentum is about $10^{48}$ g $cm^{2}s^{-1}$(Pijpers \cite{Pijpers};
    Antia et al. \cite{Antia}; Komm et al. \cite{Komm}). Thus, the mean angular momentum
    loss rate is not less than $10^{40}$ g $cm^{2}s^{-1}yr^{-1}$. Using the formula for
    angular momentum loss (Mestel \cite{Mestel84}; Kawaler \cite{Kawaler88}),
    \begin{equation}
      \frac{dJ}{dt}=\frac{2}{3}\frac{dM}{dt}R^{2}\Omega_{s}(\frac{r_{A}}{R})^{2}\,,
    \end{equation}
    and adopting the values of $\frac{dJ}{dt}$, $\frac{dM}{dt}$, and $\Omega_{s}$
    as $10^{40}$ g $cm^{2}s^{-1}yr^{-1}$, $10^{-14}$ $M_{\odot}yr^{-1}$,
    and $10^{-5}$ rad $s^{-1}$, respectively, we can get the
    corotational radius $r_{A}$ of solar wind is 100$R_{\odot}$.
    It seems difficult for the solar wind to remain corotation with the Sun
    at a distance of 100$R_{\odot}$. In this paper, we adopt a low
    initial angular momentum and ignore solar magnetic braking at
    the surface. This situation was also studied by Eggenberger et al.
    (\cite{Eggenberger}). Circulation is
    ignored because the circulation is very weak in a no-wind star,
    or it can even vanish in slow rotators (Zahn \cite{Zahn92}). In our models,
    the total angular momentum is conserved, and the secular contraction and/or
    expansion are dominant to induce interior differential rotation.

\subsection{Magnetic effects}
    Spruit (\cite{Spruit99}) reviewed the known magnetic instabilities in
    differentially rotating, stably stratified stellar interiors. Recently,
    Spruit (\cite{Spruit99}; \cite{Spruit02}) and Maeder \& Meynet
    (\cite{Maeder03}; \cite{Maeder04}) developed the Tayler-Spruit dynamo
    theory, which can generate magnetic fields in the radiative
    interior of differentially rotating stars. These fields are
    predominantly azimuthal components, $B\sim B_{\phi}$. The existence
    of magnetic field and magnetic instabilities in the stellar interior
    provides a process in which angular momentum and chemical compositions can be
    transported by magnetic stresses and magnetic instabilities. In this paper,
    we have assumed that the transport of angular momentum and mixing
    of material can be treated as a diffusion process. In the following,
    we give the diffusion coefficient for this process.

    In the plasma, the magnetic Prandtl number is (Brandenburg \&
    Subramanian \cite{Brandenburg})
    \begin{equation}
       P_{m}\equiv\frac{\nu}{\eta}=
       1.1\times10^{-4}(\frac{T}{10^{6}})^{4}(\frac{\rho}{0.1 g cm^{-3}})^{-1}
       (\frac{\ln \Lambda}{20})^{-2}\,,
    \end{equation}
    where $\nu$ is the kinematic viscosity, $\eta$ the
    magnetic diffusivity, and the $\ln \Lambda$ is the coulomb logarithm.
    In the solar radiative region ($T\sim 8\times10^{6}-2\times10^{6}$ $K$,
    $\rho \sim 20-0.2$ g$cm^{-3}$), the $P_{m}$ is about
    $10^{-2}-10^{-3}$. The $P_{m}$ is about $10^{-4}-10^{-7}$ in the
    solar convective zone (Brandenburg \& Subramanian \cite{Brandenburg}).
    We assume $\ln \Lambda =10$ in the above calculation.
    Thus, we ignore viscosity in the following equation.
    For a constant magnetic diffusivity and shellular rotation, under
    axisymmetry and only considering Lorentz force, the
    azimuthal components of the induction and momentum equations are
    \begin{equation}
       \frac{\partial B_{\phi}}{\partial t}+\eta(\frac{1}{r^{2}sin^{2}\theta}
        -\nabla^{2})B_{\phi}=rsin\theta B_{r}\frac{\partial \Omega}{\partial r}
        \,,
       \label{Induc}
    \end{equation}
    \begin{equation}
       \frac{\partial \Omega}{\partial t}=\frac{1}{4\pi \rho r^{2}
         sin^{2}\theta}{\bf B}_{p}\cdot \nabla (rsin\theta B_{\phi}) \,.
       \label{Omega}
    \end{equation}
    If we assume that the effect of the magnetic diffusivity is to limit
    the growth of the toroidal field after some time (Mestel et al.
    \cite{Mestel87}), the growth of the instability
    is halted by dissipative processes that operate on a timescale $\tau$,
    so that the second term on the left-hand side of Eq. (\ref{Induc})
    may be replaced simply by $B_{\phi}/\tau$ (Barnes et al. \cite{Barnes}).
    Substituting for the second term of Eq. (\ref{Induc}) and differentiating the
    Eq. (\ref{Induc}) with respect to time, we obtain (Barnes et al. \cite{Barnes})
    \begin{equation}
       \frac{\partial^{2}B_{\phi}}{\partial^{2}t}+\frac{1}{\tau}
       \frac{\partial B_{\phi}}{\partial t}=r sin\theta {\bf B}_{r}\cdot
                              \nabla\frac{\partial \Omega}{\partial t}\,.
    \end{equation}
    For much longer times than the timescale of the instability, one would
    expect the term involving the first time derivative to dominate, so
    that (Barnes et al. \cite{Barnes})
    \begin{equation}
    \begin{array}{lll}
      \frac{\partial (rsin\theta B_{\phi})}{\partial t}&\simeq&
      \tau (rsin\theta )^{2}{\bf B}_{r}\cdot\nabla[\frac{1}{4\pi\rho
      (rsin\theta)^{2}}{\bf B}_{p}\\
      & &\cdot\nabla (rsin\theta B_{\phi})]   \\
      &\approx&\nabla\cdot[D_{m}\nabla (rsin\theta B_{\phi})]\,.
    \end{array}
    \label{Diffu}
    \end{equation}
    Equation (\ref{Diffu}) is an approximate diffusion equation for
    $rsin\theta B_{\phi}$ with a diffusion coefficient
    \begin{equation}
       D_{m}=\frac{\tau B_{r}^{2}}{4\pi \rho}\,.
    \end{equation}

    For a steady equilibrium, the dissipating timescale has to match the
    growth timescale of the instability. In rotating stars under the
    condition $\omega_{A}\ll \Omega$, the growth time scale of the magnetic
    instability is (Pitts \& Tayler \cite{Pitts}; Spruit \cite{Spruit99})
    \begin{equation}
      \sigma^{-1}=\frac{\Omega}{\omega_{A}^{2}} \,,
    \end{equation}
    where
    \begin{equation}
      \omega_{A}=\frac{B}{(4\pi \rho)^{1/2}r}\,,
    \end{equation}
    is the Alfv$\acute{e}$n frequency. If $\tau$= $\sigma^{-1}$, one can
    get
    \begin{equation}
    \begin{array}{lll}
       D_{m}&=&\frac{B_{r}^{2}}{4\pi \rho}\frac{\Omega}{\omega_{A}^{2}}\\
       &=&r^{2}\Omega \frac{B_{r}^{2}}{B^{2}}\,.
    \end{array}
    \label{Coef}
    \end{equation}

    Note the diffusion coefficient $D_{m}$ derived from the induction and
    momentum equations, so that one would expect that the angular momentum
    transport and the mixing of elements caused by magnetic fields
    obey a similar diffusion equation, such as Eqs. (\ref{diffu1}) and
    (\ref{diffu2}) with the diffusion coefficient $D_{m}$. Although
    this diffusion coefficient is used in the equations of angular
    momentum transport and the mixing of elements, which is only an
    assumption, expression (\ref{Coef}) hints that this diffusion coefficient
    is related to the efficiency of the angular momentum transport.
    Expression (\ref{Coef}) can be rewritten as
    \begin{equation}
       D_{m}=r^{2}\Omega \frac{v^{2}_{rA}}{v^{2}_{A}}            \, ,
    \end{equation}
    where
    \begin{equation}
       v^{2}_{rA}=\frac{B^{2}_{r}}{4\pi \rho}            \, ,
    \end{equation}
    \begin{equation}
       v^{2}_{A}=\frac{B^{2}}{4\pi \rho}            \, ,
    \end{equation}
    is the Alfv$\acute{e}$n velocity square. If the angular momentum is
    transported by an Alfv$\acute{e}$n wave, and the magnetic field B
    dominates as an azimuthal field,
    so that the larger the radio $v^{2}_{rA}$/$v^{2}_{A}$, the more
    efficient the angular momentum transport.

\section{ Numerical calculation and results}

\subsection{Choice of parameters }
   The code originally written by Paczy$\acute{n}$ski (\cite{Paczynski69};
   \cite{Paczynski70}) was updated by Sienkiewicz in 1995 and Yang et
   al. (\cite{Yang}). We modified it to incorporate the hydrostatic effects
   of rotation on the equations of stellar structure, using the method of
   Kippenhahn-Meynet (Kippenhahn \& Thomas \cite{Kippenhahn70};
   Meynet \& Maeder \cite{Meynet}). The models are calculated
   using the OPAL equation of state (Rogers et al. \cite{Rogers}),
   OPAL opacity (Iglesias \& Rogers \cite{Iglesias}), and
   the Alexander \& Ferguson (\cite{Alexander}) opacity table for
   low temperature. Element diffusion is incorporated for helium
   and metals (Thoul et al. \cite{Thoul}). The nuclear reaction
   rates have been updated according to Bahcall \&
   Pinsonneault (\cite{Bahcall}). Energy transfer by convection is treated
   according to the standard mixing length theory, and the boundaries
   of the convection zones are determined by the Schwarzschild criterion.
   We adopt the solar age as 4.6$\times 10^{9}$ year, luminosity
   $L_{\odot}$=3.844$\times10^{33}$ erg$s^{-1}$, radius
   $R_{\odot}$=6.96$\times 10^{10}$ cm, and the ratio of heavy
   elements to hydrogen by mass $Z/X$=0.023 (Grevesse \& Sauval
   \cite{Grevesse}).

   As mentioned above, we assume that solid body rotation was enforced in
   the convective region. This assumption has been used by Pinsonneault et al
   (\cite{Pinsonneault}), Chaboyer et al (\cite{Chaboyer95}), and
   Huang (\cite{Huang04b}). The total initial angular momentum is a
   free parameter that is adjusted until the surface velocities of the
   solar-age models are near the solar surface velocity. We take the
   rotation rate of the Sun as a solid-body rotation, about
   2.72 $\times10^{-6}$ rad/s  (Komm et al. \cite{Komm}), as the rotation rate of
   convective zone of our model.

   The strength and spatiotemporal distribution of magnetic fields inside
   the star are poorly known. Classical dynamo models predict toroidal
   fields that are not stronger than about $10^{4}$ Gauss. But Choudhuri
   \& Gilman (\cite{Choudhuri}), D'Silva \& Choudhuri (\cite{D'Silva}),
    and Caligari et al. (\cite{Caligari}) have pointed out the value
    of the magnetic field at the bottom of convective zone as around
    $10^{5}$ Gauss. The virial theorem (Parker \cite{Parker}) sets
    an upper limit to the magnitude of the average
   solar magnetic field: $\langle B\rangle \leq 10^{8}$ Gauss. Dudorov et al.
   (\cite{Dudorov}) estimated the value of the poloidal magnetic field in the
   solar radiative zone to be about the order of unity. Using a kinematic
   model with prescribed internal rotation and a standard solar model, Fox and
   Bernstein (\cite{Fox}) investigated the existence of large-scale magnetic
   fields in the Sun, and found that the ratio of poloidal to toroidal
   components of the magnetic field is about $10^{-10}$.
   Although the ratio of $B_{r}$ to $B$ may change with the
   radial coordinate $r$, the rotation rate $\Omega$, and the time
   $t$, the relation of $\frac{B_{r}}{B}$ to $r$, $\Omega$, and $t$ is
   unclear. As a first test, we take $B_{r}/B$ to be a constant, and take the
   order of the ratio of $B_{r}$ to $B$ as $10^{-5}$ in the radiative region.

   In order to study the effects of magnetic fields, we
   construct three different types of solar models. These models are
   labelled as follows:
    \begin{itemize}
      \item M1: Model with no rotation and no magnetic field but with
       element diffusion;
      \item M2: Model with rotation and no magnetic field but with element
       diffusion;
      \item M3: Model with rotation and a magnetic field and element
       diffusion, but without including the effects of secular shear
       instability.
    \end{itemize}
   Some parameters are summarized in Table~\ref{Parat}. The mixing-length
   parameters $\alpha$, $Z_{0}$, $X_{0}$, $f_{\Omega}$, and $f_{c}$ are free
   parameters. The $X_{0}$, $Z_{0}$ and $\alpha$ are adjusted until the solar-age
   model has the values of the solar luminosity, radius, and $(Z/X)_{s}$.
   Then $Y_{0}$ is determined by $Y_{0}=1-X_{0}-Z_{0}$. The parameter
   $f_{\Omega}$ was adjusted to fit the radial profiles of the angular
   velocity, which were got from helioseismology in the solar interior.
   Parameter $f_{c}$ was adjusted to get the best radial profile of the
   sound speed. Finally, $Y_{s}$, $(Z/X)_{s}$ and $R_{bcz}$ are the results
   of calculation. All models are constructed by evolving a fully
   convective, pre-main-sequence, one solar mass model to the age of
   the present Sun.

%------------------------------------------------------
   \begin{table}
   \caption[]{Model parameters}
   \label{Parat}
   \centering
   \begin{tabular}{c c c c}
   \hline\hline
     parameter & M1 &  M2 & M3 \\
   \hline
     $\alpha$            & 1.797   & 1.790   & 1.790     \\
     $Y_{0}$             &0.2708   & 0.2707  & 0.2707    \\
     $(Z/X)_{0}$         &0.0276   & 0.0277  & 0.0277    \\
     $Y_{s}$             & 0.2437  &0.2439   & 0.2441   \\
     $(Z/X)_{s}$         &0.023    & 0.0231   &0.0232  \\
     $R_{bcz}/R_{\odot}$ &0.709    &0.710    & 0.711 \\
     $f_{\Omega}$        &         &1.0      &6.5$\times10^{-3}$\\
     $f_{c}$             &         &0.02     &0.03\\
   \hline
   \end{tabular}
   \begin{list}{}{}
     \item Note.--$\alpha$ is the mixing-length parameter;
       $Y_{0}$ and $(Z/X)_{0}$ are the initial chemical compositions;
       $Y_{s}$ and $(Z/X)_{s}$ are the surface compositions at an
       age of 4.6 $\times10^{9}$ years; $R_{bcz}$ is the radius at
       the base of the convective zone.
   \end{list}
   \end{table}
%---------------------------------------------------------------

\subsection{Transport of angular momentum }

   We define the angular momentum density J(r, t) and total angular
   momentum A(t) as
   \begin{equation}
     J(r, t)=\frac{8\pi}{3}\rho r^{4}\Omega(r, t) \,,
   \end{equation}
   \begin{equation}
     A(t)=\int^{R_{\odot}}_{0}J(r, t)dr\,,
   \end{equation}
   where $t$ is time.
%-----------------------------------------------------------
   \begin{figure*}
     \includegraphics[angle=-90, width=16cm]{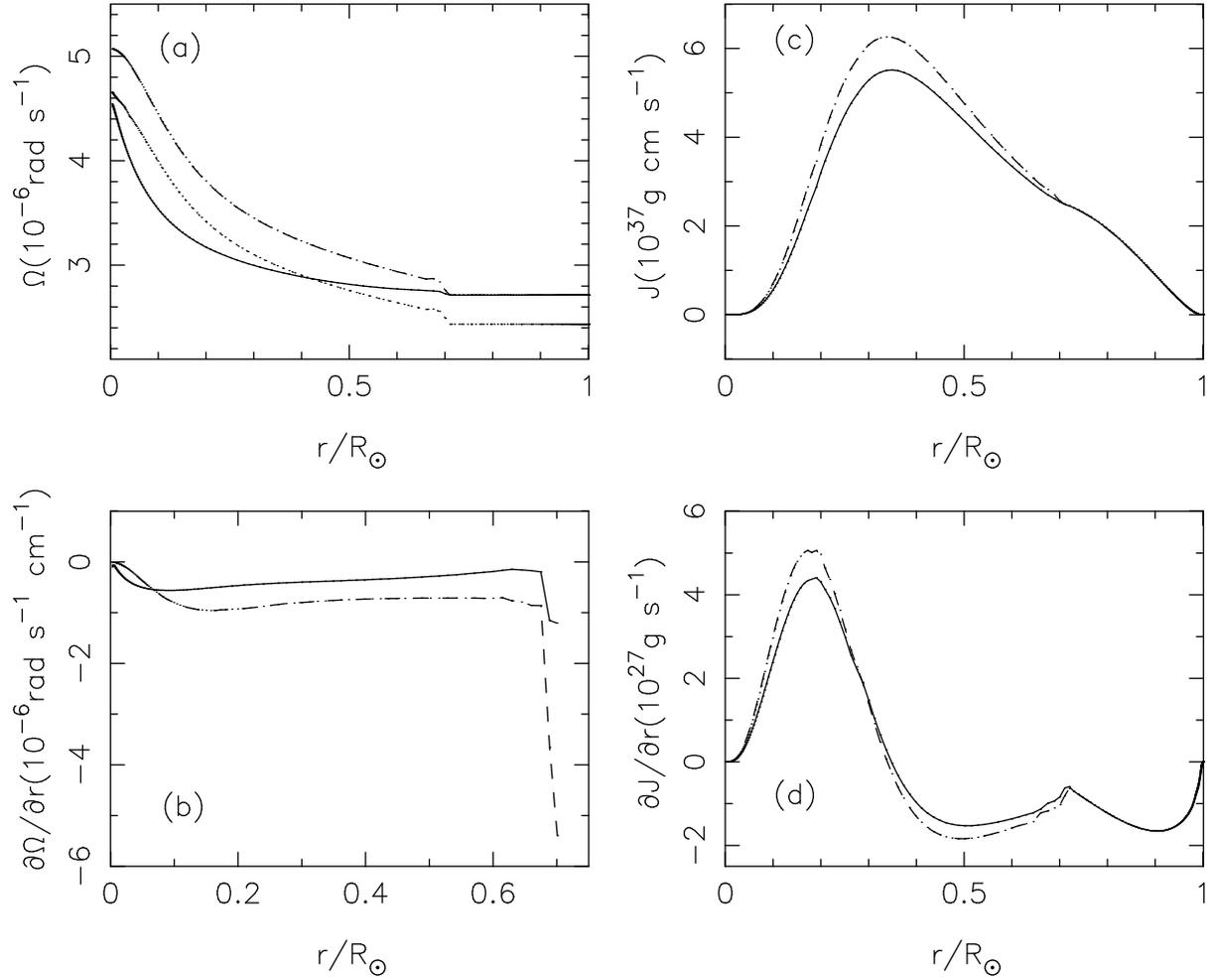}
     \centering
       \caption{Angular velocity $\Omega$, angular momentum density J, and their
       differentiation with respect to radius as a function of radius in the
       present-day Sun. (a) The solid line is model M3 with total initial angular
       momentum $A=1.91\times10^{48}$ g $cm^{2}s^{-1}$. The long-dashed line
       and the dotted line show model M2 with $A=2.1\times10^{48}$ g
       $cm^{2}s^{-1}$ and $A=1.91\times10^{48}$ g $cm^{2}s^{-1}$, respectively.
       (b, c, and d) The solid line shows model M3 with $A=1.91\times10^{48}$ g
       $cm^{2}s^{-1}$, and the long-dashed line is M2 with $A=2.1\times10^{48}$ g
       $cm^{2} s^{-1}$.
        }
       \label{omj}
   \end{figure*}
%-----------------------------------------------------------

   In Figs.~\ref{omj}a and \ref{omj}b, we present the rotation rates and their
   differentiation with respect to radius $r$ as a function $r$ at the age
   of 4.6 G years. The radial profile of the rotation rate of model M3 is flatter
   than of model M2. In our models, the radial gradient of angular
   velocity is negative. But the helioseismology has revealed that the gradient
   is positive at a low latitude and negative at a high latitude at the bottom of
   the solar convective zone (Schou et al. \cite{Schou}).
   At around $r$=0.7 $R_{\odot}$ in models M2 and M3,
   a sharp radial change in the rotation curve takes place. The thickness
   of the layer of sharp radial change in the rotation rate is about
   0.036 $R_{\odot}$. Helioseismoloy also shows that that layer exists,
   and the thickness of the layer is around between 0.02 $R_{\odot}$ and 0.05
   $R_{\odot}$ (Christensen-Dalsgaard et al. \cite{Christensen91};
   Kosovichev \cite{Kosovichev}; Basu \& Antia \cite{Basu97a};
   Basu \cite{Basu97b}; Corbard et al. \cite{Corbard98}; Corbard et
   al. \cite{Corbard99}; Charbonneau et al. \cite{Charbonneau99}). The rotation
   rate of model M3 is lower for $r<$0.4 $R_{\odot}$ and higher for
   $r>$0.4 $R_{\odot}$ than for model M2, which has the same total angular
   momentum as model M3. When M2 and M3 have the same surface rotation
   rate, the rotation rate of M2 is higher than it is for M3 in whole
   radiative region.

   The diffusion coefficient of hydrodynamical model $D_{d}$ is very small
   in the inner portion of the radiative region. It is ineffective to transport
   angular momentum outwards. But the diffusion coefficient of magnetic model
   $D_{m}$ is large enough to transport angular momentum from the core to
   the convective zone. That is to say magnetic fields are more efficient than
   the secular shear instabilities in transport angular momentum. The magnetic
   fields not only transport angular momentum outwards but strengthen the
   coupling between the radiative and convective zone. Thus, under with the same
   total angular momentum, the rotation rate of model M3 is lower for
   $r<$0.4 $R_{\odot}$ but is higher for $r>$0.4 $R_{\odot}$ than the one
   for model M2 (Fig.\ref{omj}a); the gradient of the rotation rate
   of model M3 is significantly reduced beneath the convective zone
   (Fig.~\ref{omj}b).

   The angular momentum density J and its differentiation with respect
   to $r$ of models M2 and M3, both with the same surface rotation rate
   at the age of 4.6 Gyr, are shown in Figs.~\ref{omj}c and \ref{omj}d.
   The total angular momentum of model M3 is $1.91\times10^{48}$
   g $cm^{2}g^{-1}$, which is consistent with helioseismic
   results (1.900 $ \pm $ 0.0015)$\times10^{48}$ g $cm^{2}s^{-1}$
   (Pijpers \cite{Pijpers}), 1.91$\times10^{48}$ g $cm^{2}s^{-1}$
   (Antia et al. \cite{Antia}), or (1.94 $ \pm $ 0.05)$\times$
   $10^{48}$ g $cm^{2}s^{-1}$ (Komm et al. \cite{Komm}), but that of
   model M2 is $2.1\times10^{48}$ g $cm^{2}g^{-1}$, which is higher
   than helioseismic results. The angular momentum density J shows a maximum
   $5.51\times10^{37}$ g cm $s^{-1}$ at $r$=0.35 $R_{\odot}$ of model M3,
   and $6.1\times10^{37}$ g cm $s^{-1}$ at $r$=0.34 $R_{\odot}$ of model M2.
   Using the rotation rates obtained by inversions of helioseismology, Komm et
   al. (\cite{Komm}) gave the maximum of J is $5.5\times10^{37}$
   g cm $s^{-1}$ for the Michelson Doppler Image (MDI) data, or
   $5.32\times10^{37}$ g cm $s^{-1}$ for the Global Oscillation Network
   Group (GONG) data near $r\approx$0.38 $R_{\odot}$.

   Model M3 is consistent with the Sun in the total solar angular
   momentum, the maximum of angular momentum density and the surface rotation
   rate. The magnetic fields are very important in the angular momentum
   transport in radiative interior. In our model, the core is rotating
   faster than the rest of the radiative interior. However, helioseismic
   results show the solar core is rotating more slowly than the
   rest of the radiative interior (Chaplin et al. \cite{Chaplin}; Elsworth et
   al. \cite{Elsworth}; Tomczyk et al. \cite{Tomczyk}), or else it is a
   solid-body rotation (Charbonneau et al. \cite{Charbonneau98}). Some mechanism
   of angular momentum transport may be missed, or the value of $B_{r}/B$
   should be larger than the one used in the core in our model.
   A gravity wave may be the candidate for angular momentum transport in
   the solar core, because the conditions for the known hydrodynamic
   instabilities occurrence are not satisfied in a core that is rotating
   slowly (Spruit et al. \cite{Spruit83}).

\subsection{Mixing and its effects on the solar structure}
%______________________________________________________________
   \begin{figure*}
     \includegraphics[angle=-90, width=16cm]{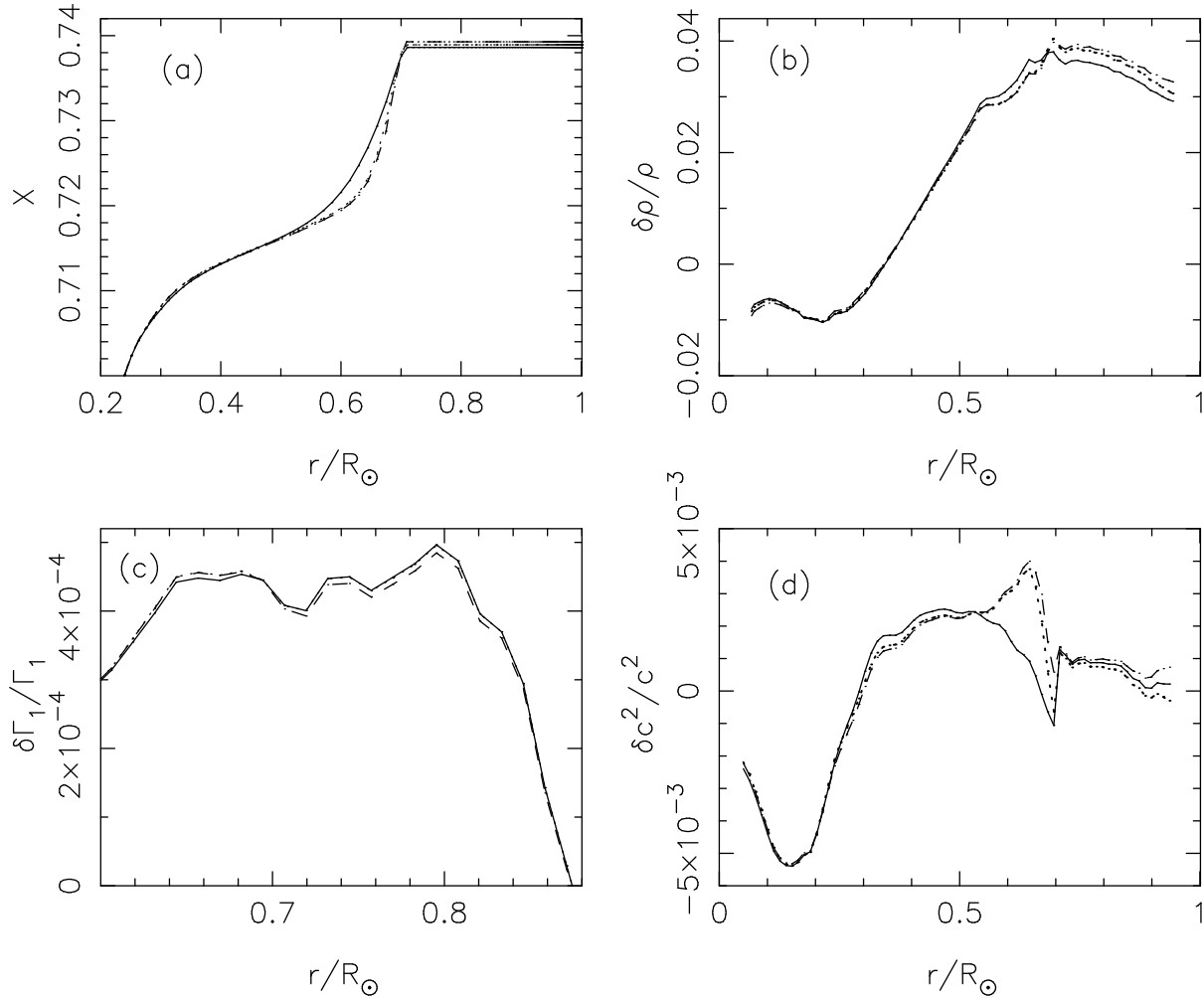}
     \centering
       \caption{(a) Radial distribution of hydrogen mass fraction. (b) Density
       difference between the Sun and the model. (c) Adiabatic index
       difference between the Sun and the model. (d) Squared sound-speed
       difference between the Sun and the model. All the differences
       are in the sense (Sun)-(Model). The solid line refers to model M3. The
       long-dashed line refers to model M1. The dotted line refers to model
       M2 with total angular momentum $2.1\times10^{48}$ g $cm^{2}g^{-1}$.
              }
       \label{xrgc}
   \end{figure*}
%______________________________________________________________

   The radial profiles of hydrogen mass fraction $X$ are plotted in
   Fig.~\ref{xrgc}a. In Figs.~\ref{xrgc}b, \ref{xrgc}c, and \ref{xrgc}d,
   we compare the density, adiabatic index, and squared sound speed of
   our models with the helioseismic inversion results given by Basu et
   al. (\cite{Basu97c}; \cite{Basu00}).
   The mixing of magnetic instabilities causes more of an increase of
   the hydrogen abundance just beneath the convective zone, but a decrease
   in the convective zone and near $r$=0.3 $R_{\odot}$, than that of model
   M1 and M2. The increase of $X$ between 0.5 $R_{\odot}$ and 0.7 $R_{\odot}$
   causes a decrease in the mean molecular weight $\mu$ and density $\rho$,
   so that the difference in density $\rho$ increases in that region
   (Fig.~\ref{xrgc}b). Since the squared sound speed $c^{2}\propto T/\mu$,
   reducing $\mu$ must lead to an increase in $c^{2}$. The difference of squared
   sound speed, in the sense ($c^{2}_{\odot}-c^{2}_{model}$)/$c^{2}_{model}$, has a
   significant decrease, from 0.004 down to 0.0009 at $r$=0.65 $R_{\odot}$
   (Fig.~\ref{xrgc}d), whereas the effect is the opposite between 0.3 $R_{\odot}$
   and 0.5 $R_{\odot}$. In Fig.~\ref{xrgc}d, there is a negative value,
   -1.06$\times10^{-3}$, of $\delta c^{2}/c^{2}$ for models M2 and M3 at the bottom
   of the convective zone. That region is where the sharp radial change in $X$
   happens due to the helium settling. The strong mixing leads to this negative
   value. The effect of mixing to the adiabatic index is very small, and the
   increase in $X$ causes a small increase in the adiabatic index (Fig.~\ref{xrgc}c).

   Although material mixing can account for the bump in the
   sound-speed's squared difference between 0.6 $R_{\odot}$ and
   0.7 $R_{\odot}$, the mass conservation causes a decrease in $X$, an
   increase in $\mu$, and hence an increase in $\delta c^{2}/c^{2}$
   between 0.3 $R_{\odot}$ and 0.5 $R_{\odot}$. This change is not
   expected, but it must occur in a mixing model.

\section{Discussion and conclusions}

   Parameter $f_{\Omega}$ is significantly less than unity in our model.
   This could be a consequence of overestimating the ratio of $B_{r}$ to
   $B$. The average ratio of the poloidal to toroidal field strength given by
   Fox \& Bernstein (\cite{Fox}) is $10^{-10}$. If we take a lower value
   for $B_{r}/B$, such as $10^{-6}$, instead of the previous one, we
   find that $f_{\Omega}$ is 1 order of magnitude. For simplicity, $f_{\Omega}$
   and $f_{c}$ were taken as constant so the assumption might not be strictly
   correct. If we take the parameter as $f_{\Omega}(r, t)$, then, the value of
   $f_{\Omega}(r, t)$ can be more than 6.5$\times10^{-3}$ in most regions and
   times. But the uncertainty of magnetic $B_{r}/B$ is larger than that of
   $f_{\Omega}$. We take $f_{\Omega}$ as a constant. This is also a reason for
   low $f_{\Omega}$. The parameter $f_{c}$ is less than unity, too. In the
   magnetic model, the angular momentum can be transported by both magnetic
   stresses and magnetohydrodynamical (MHD) instabilities, but only the MHD
   instabilities can mix material. So, the efficiency of material mixing must
   be less than that of angular momentum transport, i.e. low $f_{c}$.

   The magnetic field is more efficient than the secular shear instabilities
   in angular momentum transport outwards. Thus the rotation
   rate of model M3 is lower for $r<$0.4 $R_{\odot}$ and higher for
   $r>$0.4 $R_{\odot}$ than is model M2, which has the same total angular
   momentum as model M3. The model with the magnetic field is
   consistent with the heliosiesmic results in the solar total
   angular momentum, the maximum of angular momentum density, and
   the surface rotation rate.

   In this paper, we obtained the diffusion coefficient of magnetic angular
   momentum transport, which depends on the radius, rotation rate, and
   magnetic fields. Using this diffusion coefficient, we find that the
   angular momentum can be efficiently redistributed by magnetic fields.
   The radio of $B_{r}$ to $B$ may be $10^{-6}$ in the solar radiative
   interior. The coupling between the radiative region and convective zone
   can be strengthened by magnetic fields, hence the sharp gradient of
   rotation rote (Fig.~\ref{omj}b) is reduced at the bottom of the convective
   zone. In our model, the thickness of the layer of the sharp radial
   change in the rotation rate is about 0.036 $R_{\odot}$, and the
   rotation is almost uniform in the radiative region. There is a maximum,
   5.51$\times 10^{37}$ g $cm^{2}g^{-1}$, of angular momentum density J
   at $r$=0.35 $R_{\odot}$ for the model with a magnetic field. Material
   mixing associated with the angular momentum transfer leads to a change
   in the distribution of the chemical compositions, so that there
   is noticeable improvement in the profiles of the sound speed and density.

-------------------------------------------

\begin{acknowledgements}
   This work was supported by the NSFC through project 10473021, the
   National Key Fundamental Research Project G2000078401, and the
   Yunnan Science Foundation Council 2003A0027R.

\end{acknowledgements}

\end{document}